\documentstyle[12pt]{article}
\newcommand{\titul}[1] {\begin{center}{\large\bf #1 }
\end{center}\vskip 1.cm}

\newcommand{\abstr}[1] {{\begin{center} \vskip .5cm {\bf Abstract
                        \vspace{0pt}} \end{center}}\begin{quote} #1
                        \end{quote}}

\begin{document}
\begin{titlepage}
\titul{
Next-to-next-to-leading order  QCD
analysis  of  combined data for  $xF_3$ structure function
and higher--twist contribution. }

\begin{center}
{\bf A.V. Sidorov}\\ [1cm]
 {\em Bogoliubov Laboratory of Theoretical Physics\\
 Joint Institute for Nuclear Research, 141980 Dubna, Russia}\\
\end{center}

\abstr{

 The simultaneous QCD analysis of the $xF_3$
structure functions measured
in deep-inelastic scattering by several collaborations
is done up to 3--loop  order of QCD.
The x dependence of the higher--twist contribution is evaluated
and turns out to be in a qualitative agreement with the results of
"old" CCFR data analysis and with renormalon approach predictions.
The Gross--Llewellyn Smith sum rule and its
higher--twist corrections are evaluated.
}

\begin{center}
{\it{Talk presented at XIII International Seminar
on High Enetgy Physics Problems
"Relativistic Nuclear Physics and Quantum Chromodynamics",
Dubna, Russia, September 1996.}}
\end{center}

\end{titlepage}

{\bf 1.} The experimental data of the CCFR collaboration
(we'll call them "old")
obtained at Fermilab Tevatron
\cite{CCFR} for the $xF_3$  structure functions of
deep-inelastic scattering of neutrinos and antineutrinos on an iron target
provide an important means of accurate comparison of QCD with experiment.
However,
in view of revision of "old" data announced dy CCFR collaboration \cite{Warsaw}
the question arises: what can we say about the comparison of the QCD
predictions on $Q^2$ dependence of the $xF_3(x,Q^2)$ structure function (SF)
based on the data of neutrino DIS experiments different from those of CCFR?

 In the present note,
a combined fit of the experimental data of the CDHS \cite{CDHS},
SCAT \cite{SCAT}, BEBC-WA59  \cite{BEBC}, BEBC--Gargameile  \cite{BG}
and JINR-IHEP \cite{JI} collaborations
for the $xF_3$  structure functions is done in order to determine
the x dependence of the SF, higher twist (HT) contribution
and the value of the scale parameter $\Lambda_{\overline{MS}}$. \\ [3mm]

{\bf 2.} We'll use, for the QCD analysis,
the Jacobi polynomial expansion method
proposed in \cite{PS}.
It was developed in \cite{PS}-\cite{nnl}
 and applied for the  3--loop  order of pQCD
to fit  $F_2$ \cite{PKK} and $xF_3$ data \cite{nnl,htnnl}.

The $Q^2$ - evolution of the moments $M_3^{pQCD}(N,Q^2)$ is
given by the well known perturbative QCD \cite{s4,s5} formula:

\begin{eqnarray}
M_3^{pQCD}(N,Q^2)
& =& \left [ \frac{\alpha _{S}\left ( Q_{0}^{2}\right )}
{\alpha _{S}\left ( Q^{2}\right )}\right ]^{d_{N}}
H_{N}\left (  Q_{0}^{2},Q^{2}\right )
M_3^{pQCD}(N,Q^2_0) ,~~~N = 2,3, ...  \label{m3q2} \\
d_N & = & \gamma^{(0),N}/2\beta_0,
. \nonumber
\end{eqnarray}

The factor $H_{N}\left (  Q_{0}^{2},Q^{2}\right )$ contains all next--
and the next--to--next--to--leading order QCD corrections
\footnote{For reviews and references on higher order
QCD results see\cite{vanNeerven}.}
and is constructed in
accordance with \cite{nnl} based on theoretical results of \cite{thnnl}.

The expression (\ref{m3q2}) provides an input for
reconstruction of the SF by the Jacobi polynomial method.
Following the method \cite{Kri,BCDMS}, we can write
 the structure
function $~xF_3~$ in the form:
\begin{equation}
xF_{3}^{pQCD}(x,Q^2)=x^{\alpha }(1-x)^{\beta}\sum_{n=0}^{N_{max}}
\Theta_n^{\alpha , \beta }
(x)\sum_{j=0}^{n}c_{j}^{(n)}{(\a ,\beta )}
M_{3}^{QCD} \left (j+2, Q^{2}\right ),   \\
\label{e7}
\end{equation}
where $~\Theta^{\alpha \beta}_{n}(x)~$ is a set of Jacobi polynomials and
$~c^{n}_{j}(\alpha,\beta)~$ are coefficients of the series of
$~\Theta^{\alpha,\beta}_{n}(x)~$ in powers of x:
\begin{equation}
\Theta_{n} ^{\a , \beta}(x)=
\sum_{j=0}^{n}c_{j}^{(n)}{(\a ,\beta )}x^j .
\label{e9}
\end{equation}

The unknown coefficients $M_3(N,Q^2_0)$ in (\ref{m3q2}) could be parametrised
as Mellin moments of some function:
\begin{eqnarray}
M_3^{pQCD}(N,Q^2_0)&=&\int_{0}^{1}dx{x^{N-2}}Ax^b(1-x)^c(1+\gamma x),
~~~ N = 2,3, ...
\label{Mellf30}
\end{eqnarray}

To extract the HT contribution, the nonsinglet SF is
parameterized as follows:
\begin{eqnarray}
xF_3(x,Q^2)=xF_3^{pQCD}(x,Q^2)+h(x)/Q^2,
\label{xf3}
\end{eqnarray}
where the $Q^2$ dependence of the first term in the r.h.s is determined by
perturbative QCD.
Constants
$h(x_i)$ (one per x--bin) parameterize the HT x dependence.
We put $x_i=$
$0.03,
~0.05,
~0.08,$ $
~0.15,
~0.25,$ $
~0.35,
~0.45,$ $
~0.50,$ $
~0.55,
~0.65,$ $
~0.80$
for $i=1,2...11$. The HT contribution for $F_2$ was determined in \cite{MV}.
The values of constants $h(x_i)$
as well as the parameters A, b, c, $\gamma$  and
scale parameter $\Lambda$       are determined by fitting the
combined  set of data of 192 experimental points of $xF_3$ in a wide
kinematic region:
$0.5~GeV^2\leq~Q^2~\leq~196~GeV^2$ and $0.03~\leq~x~\leq~0.80$
 and  $Q^2_0=10~GeV^2$. We have put the number of flavors to equal 4.
In accordance with the result of \cite{CDHS} concerning  the disagreement of
their data with perturbative QCD at small x, a cut $x\geq0.35$ was used for
CDHS data.
The  TMC are taken into account to the order  $o(M_{nucl}^4/Q^4)$ .

The  nuclear
effect of the relativistic  Fermi motion is estimated
>from below by the ratio $R_F^{D/N}=F_3^D/F_3^N$ obtained in the covariant
approach in light-cone variables \cite{BrTo}.

\begin{center}
\begin{tabular}{||c|c|c|c||} \hline \hline
                                &      LO              &
       NLO          &         NNLO               \\ \hline
     $\chi^2_{d.f.}$              &  312/176            &
        316/176    &                 312/176             \\
        A                         &   6.68 $\pm$      0.38
          &  6.92  $\pm$      1.43    &  7.11   $\pm$       0.38   \\
        b                         &  0.760 $\pm$      0.027  & 0.768
         $\pm$      0.072    &  0.778 $\pm$       0.027  \\
        c                         &   4.03 $\pm$      0.07   & 3.97
          $\pm$      0.17      &  3.82 $\pm$       0.07   \\
    $\gamma$                      &  0.675 $\pm$      0.156
     &  0.452 $\pm$      0.624   &  0.189  $\pm$        0.128 \\
$\Lambda_{\overline{MS}}$ $[MeV]$ &  191   $\pm$      46
  &159   $\pm$      39       &   163  $\pm$         31   \\  \hline \hline
        $x_i$                     &\multicolumn{3}{c||}{
         $h(x_i)~[GeV^2]$ }                                                     \\  \hline
    0.03  &   0.086 $\pm$      0.087 &    0.090  $\pm$     0.091 &
        0.067 $\pm$    0.085                            \\
    0.05  &   0.001 $\pm$      0.028 &    0.022  $\pm$     0.032 &
       0.093 $\pm$      0.047                          \\
    0.08  &  -0.127 $\pm$      0.123 &    -0.094 $\pm$     0.126 &
       -0.011 $\pm$    0.131                            \\
    0.15  &  -0.286 $\pm$      0.046 &   -0.230  $\pm$     0.050 &
       -0.200 $\pm$      0.050                          \\
    0.25  &  -0.401 $\pm$      0.058 &   -0.334  $\pm$     0.056 &
       -0.327 $\pm$      0.054                          \\
    0.35  &  -0.284 $\pm$      0.073 &   -0.220  $\pm$     0.068 &
      -0.178 $\pm$      0.062                          \\
    0.45  &  -0.436 $\pm$      0.093 &   -0.366  $\pm$     0.090 &
      -0.403 $\pm$      0.083                          \\
    0.50  &   0.005 $\pm$      0.079  &   0.047  $\pm$     0.077  &
       0.036 $\pm$     0.074                           \\
    0.55  &  -0.243 $\pm$      0.069 &   -0.200  $\pm$     0.068 &
       -0.242 $\pm$      0.064                          \\
    0.65  &   0.176 $\pm$      0.063 &    0.202  $\pm$     0.072 &
       0.154 $\pm$      0.060                          \\
    0.80  &   0.020 $\pm$      0.037 &    0.024  $\pm$     0.039 &
       -0.012 $\pm$     0.039                            \\   \hline  \hline
\multicolumn{4}{p{12cm}}{{\bf Table I.} Results of the 1-, 2- ($N_{Max}=10$)
and 3- order  ($N_{Max}=8$)
QCD fit (with TMC)
of the combined $xF_3$ SF data for  $f=4$, $Q^2>0.5 GeV^2$ with
the corresponding statistical errors, normalization coefficients
and values of the HT contribution $h(x_i)$.
} \\ [3mm]
\end{tabular}
\end{center}

{\bf 3.}
Results of the fit are presented in Table 1 and Figures 1-3. The theoretical
prediction for $h(x)$  from \cite{webber} is presented  Figure 3.

The experimental values of $xF_3$ for each collaboration
were multiplied by the normalization factors
$C^{coll}$ which were considered as free parameters.
Their values are not sensitive to the order of pQCD in use and was
found to be equal to:
$C^{ BEBC-WA59}=0.92\pm 0.03$,~
$C^{ SCAT}=1.06\pm 0.03$,~
$C^{ JINR-IHEP}=1.02\pm 0.05$ and
$C^{ BEBC-Garg.}=0.97\pm 0.04$. The value of $C^{CDHS}=1$ was fixed.

The obtained value of $\Lambda_{\overline{MS}}$ is larger than that
given by a
similar analysis of CCFR data \cite{htnnl}
$\Lambda_{\overline{MS}}=134\pm57~MeV$ but exhibits
relatively small statistical errors. Results of the  NLO and NNLO fit give
the constant of strong  interaction
$\alpha_S^{NLO} (M_Z^2)=0.105\pm 0.004$ and
$\alpha_S^{NNLO} (M_Z^2)=0.107\pm 0.003$
 in agreement, within the errors,
with usual DIS results \cite{Beth95}.
Additional uncertainties  to the value of $\alpha_S (M_Z^2)$ due
to extrapolation of the $Q^2$ dependence of the SF with four flavors (f=4)
in a wide kinematic interval
$0.5~GeV^2\leq~Q^2~\leq~196~GeV^2$  were found to be 0.001 in \cite{SMS} .

The value of the perturbative part of the GLS
sum rule \cite{gls} at $Q^2=10~GeV^2$ estimated by using results of Table~1
is equal to
$\int_0^1 \frac{xF_3^{pQCD}(x)}{x}dx~=~2.60\pm0.23$ in agreement with
results of the "old" CCFR data analysis \cite{glsccfr,KaSi}.

The shape of h(x) is in
qualitative agreement with theoretical predictions of the dispersion method
of the renormalon approach \cite{webber}
( for reviews and references see \cite{RevHT}) and with
 results of the QCD analysis of "old" CCFR data presented in \cite{htnnl}.
They  obviously differ from the precise values of h(x)
for singlet ST $F_2$ presented in \cite{MV}.

Based on the results of Table~1, one can estimate
the value of the  first moment of h(x)
which contributes to the GLS sum rule \cite{gls}:
$h_1=\int_0^1 \frac{h(x)}{x}dx~.$ The obtained values:
$h_1^{LO}=-0.42\pm0.27$~
\footnote{Hereafter present the value of h(x)
in $[GeV^2]$.}
, $h_1^{NLO}=-0.29\pm0.28$
and $h_1^{NNLO}=-0.26\pm0.27$ are in agreement with theoretical
predictions of
\cite{BK} $h_1=-0.29 \pm 0.14$
and \cite{HT} $h_1=-0.47\pm0.04$ as well as
with the recent result of \cite{BHT}.      \\ [3mm]

{\bf 4.} In conclusion it should be stressed that combined fit provides
still a more precise determination of $\Lambda_{\overline{MS}}$
and $h(x_i)$ in comparison to
the analysis of "old" CCFR data \cite{htnnl}, while the shape of the SF
ruled by
parameters A, b, c and $\gamma$  is determined less accurate.
The most discrepancy with the "old"  CCFR data analysis takes place
for the HT contribution to the GLS sum rule and for the
HT x dependence at large x.

For a more precise determination of the HT
contribution to SF, the role of the nuclear effect should be clarified and
a more realistic approximation for $R_F^{Fe/N}=F_3^{Fe}/F_3^N$
is needed. We also did not take into account the
threshold effects on $Q^2$ evolution of SF due to
heavy quarks \cite{match} which
is necessary owing to a wide kinematic region
of data under consideration and could be realized
based on the mass-dependent MOM-scheme \cite{SMS}. \\ [4mm]

{\bf Acknowledgements.} \\[2mm]

The author is grateful to  S.A.~Bunyatov, A.L.~Kataev, V.G.~Krivokhizhin,
S.A.~Larin, S.V.~Mikhailov and  M.V.~Tokarev for discussions.
This investigation has been
supported in part by INTAS grant No 93-1180
and by the Russian Foundation for Fundamental
Research (RFFR) N 95-02-04314a.

\newpage

Figure captions. \\[5mm]

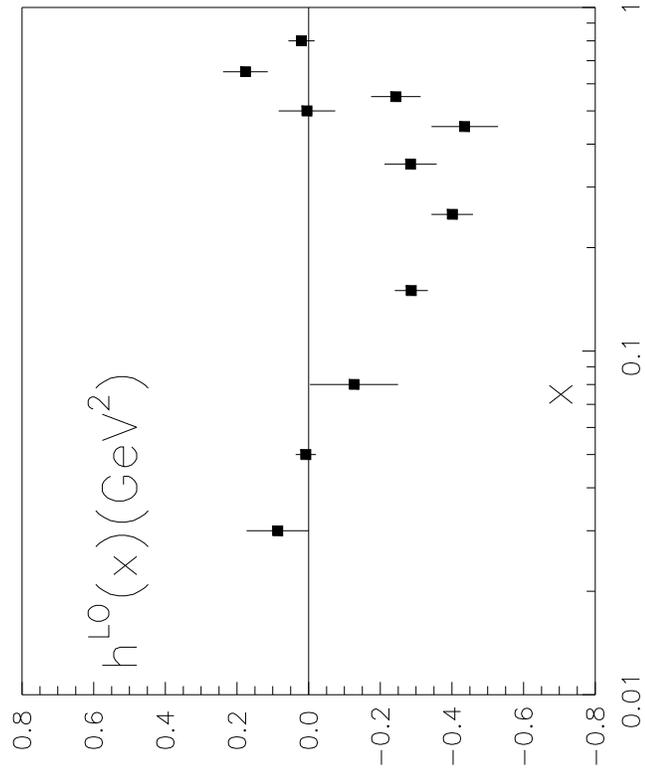
Fig.1.  Higher--twist contributions from LO fit.   \\[5mm]

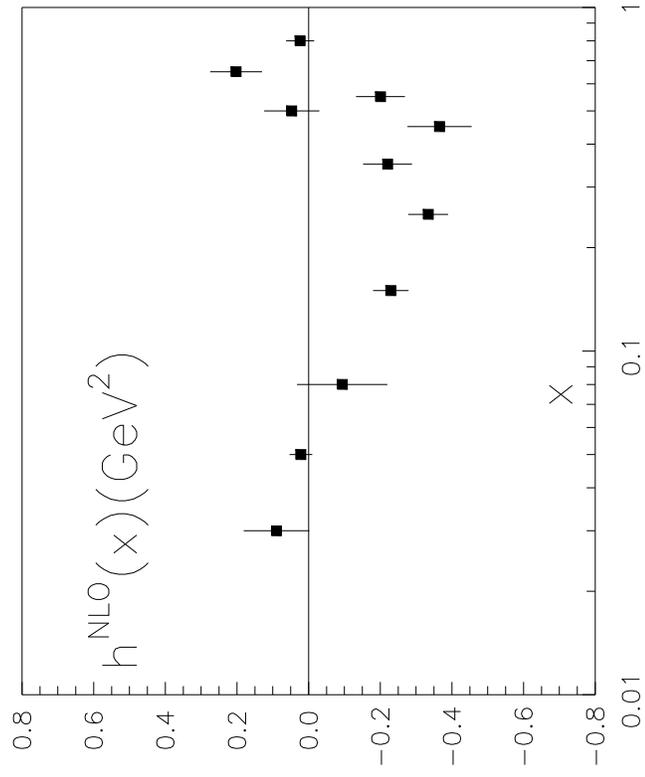
Fig.2.  Higher--twist contributions from NLO fit.  \\[5mm]

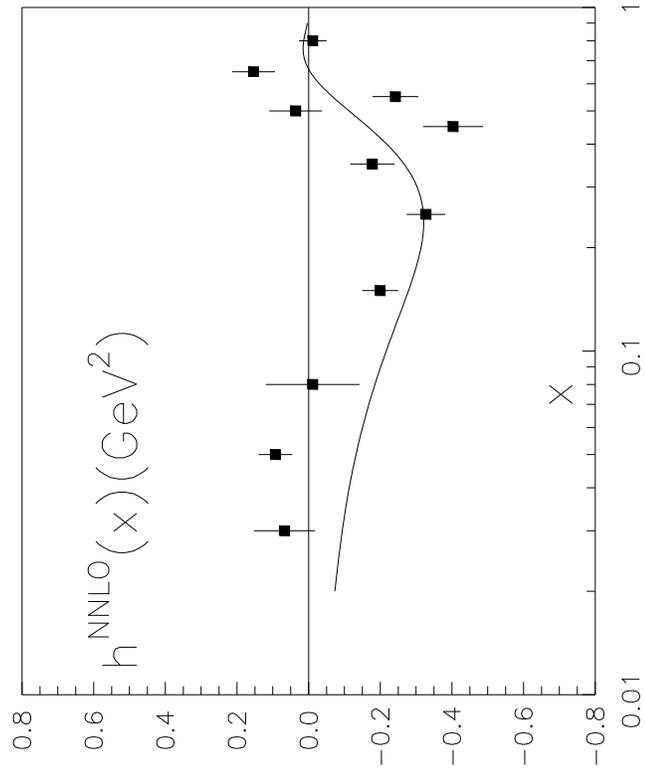
Fig.3.  Higher--twist contributions from NNLO fit and the
theoretical prediction for $h(x)$ from \cite{webber}. \\[5mm]

\end{document}